# TUTORIAL ON NEUTRON PHYSICS IN DOSIMETRY

S. Pomp[1,*]

[1] Department of physics and astronomy, Uppsala University, Box 516, 751 20 Uppsala, Sweden.

[*] Corresponding author. *E-mail address*: Stephan.Pomp@physics.uu.se (S.Pomp)

**Abstract:**

Almost since the time of the discovery of the neutron more than 70 years ago, efforts have been made to understand the effects of neutron radiation on tissue and, eventually, to use neutrons for cancer treatment. In contrast to charged particle or photon radiations which directly lead to release of electrons, neutrons interact with the nucleus and induce emission of several different types of charged particles such as protons, alpha particles or heavier ions. Therefore, a fundamental understanding of the neutron-nucleus interaction is necessary for dose calculations and treatment planning with the needed accuracy. We will discuss the concepts of dose and kerma, neutron-nucleus interactions and have a brief look at nuclear data needs and experimental facilities and set-ups where such data are measured.

**Introduction**

Dosimetry is concerned with the ability to determine the absorbed dose in matter and tissue resulting from exposure to directly and indirectly ionizing radiation. The absorbed dose is a measure of the energy deposited per unit mass in the medium by ionizing radiation and is measured in Gray, Gy, where 1 Gy = 1 J/kg. Radiobiology then uses information about dose to assess the risks and gains. A risk is increased probability to develop cancer due to exposure to a certain dose. A gain is exposure of a cancer tumour to a certain dose in order to cure it. To achieve accurate and efficient dose determination experimental data are necessary. Examples are information on the actual radiation field, nuclear cross sections, detector response, etc.

In this paper we shall focus on the role of the neutron in dosimetry. For about 70 years, efforts have been made to understand the effects of neutron radiation on tissue. The progress that has been made is enormous and it is now well known that neutrons can be used to treat certain types of cancer (Tubina et al., 1990). However, the accurate determination of the dose involved in a treatment remains difficult (Chadwick et al., 1997; ICRU Report 63, 2000). Unlike protons, neutrons carry no charge and do not directly interact with electrons. Hence the world of a neutron is in sharp contrast to the world of a proton. Protons "see" a sea of electrons with which they continuously interact. Therefore, the Bethe-Bloch equation is largely sufficient to calculate the absorbed dose delivered by protons. For 100 MeV protons stopping in tissue, nuclear reactions matter only on the level of a few percent. For a neutron, on the other hand, matter is empty space with small concentrated blobs of nuclear matter spread around. A neutron passing through a volume containing a number of nuclei has only two options; either it passes undisturbed or it interacts with a nucleus. The intensity *I* of neutrons after traveling along path *x* through matter is given by $I(x) = I_0 \cdot e^{-N \cdot \sigma \cdot x}$, where $I_0$ is the in initial intensity, *N* is the number of nuclei per unit volume and $\sigma$ is the reaction cross section. This principal behavior is similar to gamma rays. However, while gamma rays are able to ionize directly, neutrons first have to produce secondary particles. In order to deal with this fact, one often uses the kerma concept which is described in the next section.

## Kerma calculation from basic nuclear cross sections

The capability of neutrons to ionize is only indirect, i.e., via creation of secondary particles. These then can interact with the atomic electrons. The concept of kerma deals with the initial production of secondary *charged* particles. The acronym kerma stands for Kinetic Energy Released in Matter and is defined as $K = \frac{dE_{tr}}{dm}$ [1 Gy = 1 J/kg], where $dE_{tr}$ is the expectation value of the sum of the initial kinetic energies of all the charged particles liberated by uncharged particles in a material of mass *dm* (ICRU Report 63, 2000). Given the fluence $\Phi$ of uncharged particles at the same point, the kerma *K* can be obtained from kerma coefficients $k_\Phi$ using $K = \Phi \cdot k_\Phi$. The kerma coefficients can either be determined in a direct way from calorimetric measurements or calculated from the underlying basic nuclear cross sections. This is discussed in more detail by Blomgren and Olsson (2003).

Partial kerma coefficient $k_\Phi$ at neutron energy $E_n$ and for a specific target material can be obtained from microscopic cross sections using

$$k_\Phi(E_n) = N \sum_j \int E \int \frac{d^2\sigma_j(E_n)}{d\Omega dE} d\Omega dE = N \sum_j \int E \frac{d\sigma_j(E_n)}{dE} dE, \qquad (1)$$

where *N* is the number of nuclei per unit mass, *E* is the energy of the secondary charged particle and $\frac{d^2\sigma_j(E_n)}{d\Omega dE}$ is the double differential cross-section for emission of charged particle *j* at the neutron energy $E_n$. Since we are dealing with discrete data values, we can replace the integrals in Eq. 1 with sums. First the energy-differential cross section $\frac{d\sigma_j}{dE_i}$ is obtained by summation over the *k* angular bins $\Delta\Theta_k$

$$\frac{d\sigma_j(E_n)}{dE_i} = \sum_k \frac{d^2\sigma_j(E_n)}{d\Omega dE} 2\pi \sin\Theta_k \Delta\Theta_k.$$

Then we obtain the kerma coefficient from

$$k_\Phi(E_n) = N \sum_j \sum_i E_i \frac{d\sigma_j(E_n)}{dE_i} \Delta E_i, \qquad (2)$$

where $E_i$ is an energy bin of width $\Delta E_i$.

Finally, kerma coefficients $k_\Phi$ for materials composed of several elements can be obtained by summation over the $k_\Phi$ for each of the elements weighting with the relative amount (ICRU Report 63, 2000; Göttsche et al., 2009).

Kerma and dose are measured in the same units but it is important to emphasize once more that kerma describes the amount of *released* energy of secondary charged particles due to the neutron radiation while dose is concerned with the amount of *absorbed* energy from these secondary particles.

The detailed information contained in the microscopic cross sections about the various types of secondary particles, their energy, range, and angular distribution is lost within one single kerma coefficient. From a treatment-planning perspective this is unsatisfactory. While kerma coefficients can be used to obtain rough estimates, detailed information about the neutron-nucleus interactions is needed for a proper calculation of absorbed dose.

**Neutron-induced nuclear reactions from low to high energies**

The best starting point when looking at nuclear reactions is to look at which reaction channels are open, i.e., energetically possible. The elastic (n,n) and the capture (n,γ) channel are always open, i.e., these reactions occur even at the lowest energies. An answer to whether a reaction can occur at a certain energy can be obtained from calculation of the so-called reaction $Q$-value (see, e.g., Krane (1988)) $Q = (m_{\text{initial}} - m_{\text{final}})c^2 = T_{\text{initial}} - T_{\text{final}}$, where $m$ and $T$ are the masses and kinetic energies of the particles and nuclei before (initial) and after (final) the reaction. If $Q > 0$, the reaction is exothermic and energy is released. In such a case the reaction (or decay) can occur at rest or, e.g., with zero incoming kinetic neutron energy. For the elastic channel the $Q$-value is, in a sense by definition, zero. If $Q < 0$, the reaction is endothermic. This means that the reaction has a threshold energy below which the reaction cannot happen. Examples of reactions with positive $Q$-value are the $^{10}$B(n,α), with $Q$ = 2.790 MeV, and $^{14}$N(n,p) reaction, with $Q$ = 0.626 MeV. Fission of $^{235}$U is a further example. For this reaction the $Q$-value is about 200 MeV depending on the exact fission channel. Figure 1 shows cross-sections as a function of incoming neutron energy for two examples of reactions with positive $Q$-value. The cross section drops with increasing velocity $v$ of the incoming neutron and the characteristic $1/v$ behavior for such exothermic reactions in the low energy domain can be clearly seen.

With increasing incoming energy more reactions channels open and, e.g., inelastic reactions become possible. Figure 2 shows the $^{12}$C(n,n') cross section as a function of the incoming neutron energy as an example. The threshold for this reaction is at about 4.44 MeV, slightly above, for kinematics reasons, the first excited state in $^{12}$C which is at 4.4391 MeV.

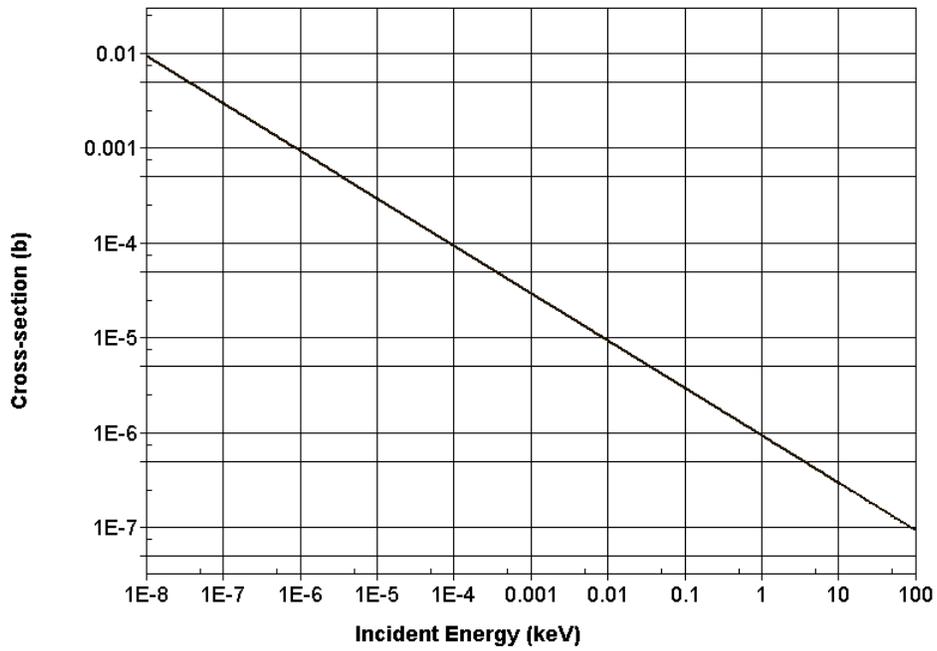

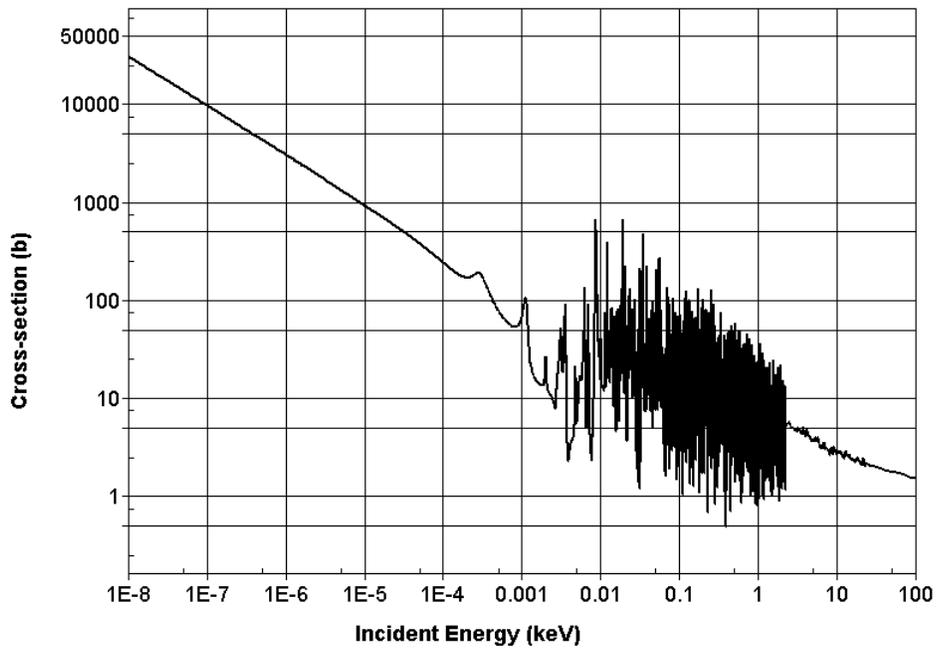

Figure 1: Cross section for the $^{16}$O(n,γ) reaction (top) and 235U(n,fission) reaction (bottom) plotted as function of the incident neutron energy. The data are from the ENDF/B-VII evaluation and have been retrieved using Janis 3.0 (Soppera et al., 2008).

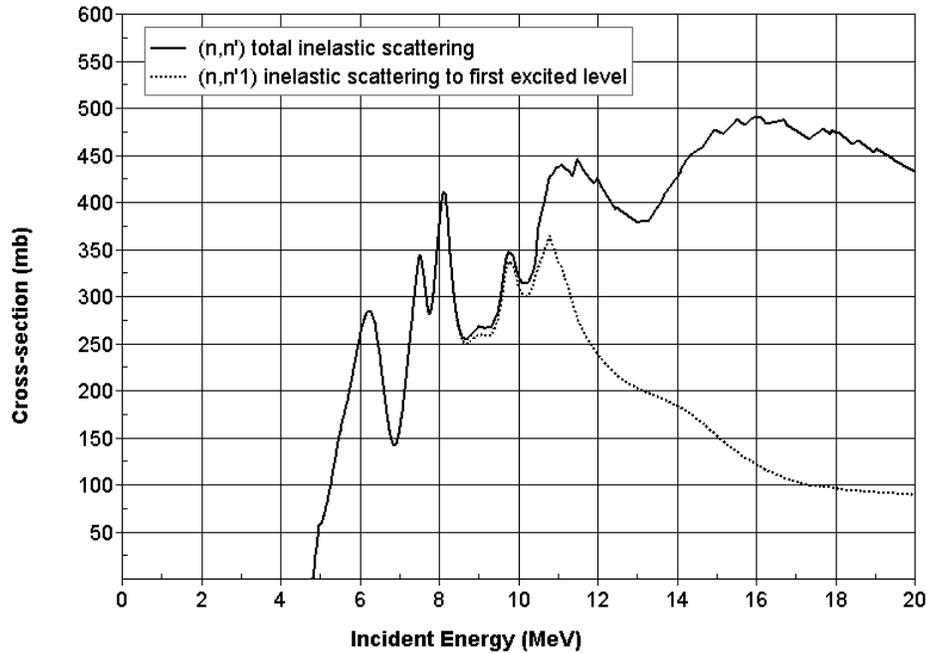

Figure 2: Cross section for the $^{12}$C(n,n') reaction plotted as function of the incident neutron energy. The lower line contains only the cross section to the first excited state in $^{12}$C. The data are from the JENDL 3.3 evaluation and have been retrieved using Janis 3.0 (Soppera et al., 2008).

Another useful concept involves the so-called neutron (proton) separation energies. It describes the necessary energy for removal of a neutron (proton) from a nucleus, usually denoted $S_n$ ($S_p$). These energies can be calculated from the difference in binding energies B (Krane, 1988):

$$S_n = B\left(^{A}_{Z}X_N\right) - B\left(^{A-1}_{Z}X_{N-1}\right).$$

Figure 3 shows a schematic view of the (n,n), (n,γ) and (n,n') reactions. The target nucleus is shown to the left and marked with (Z,N), indicating the number of protons and neutrons in the nucleus. Together with the incoming neutron the so-called compound nucleus (Z,N+1) is formed. Within the compound reaction model the final product of a nuclear reaction is assumed to be independent of the means of formation of the compound nucleus. It turns out that this model works very well at energies of a few tens of MeV. The compound state in Fig. 3 can either decay back elastically to the (Z,N) ground state by emission of a neutron with the same energy as the originally incoming neutron had. It can also decay to states below the neutron separation energy $S_n$ by emission of γ's. This is the capture reaction. If the incoming neutron energy is sufficiently high, the compound nucleus may also emit a neutron n' with a kinetic energy that is lower than the incoming neutron and reach an exited state in the (Z,N) nucleus. This is the inelastic (n,n') reaction. For completeness it should be noted that also (n,γn) reactions are possible. These involve a γ transition in the continuum above $S_n$ prior to re-emission of a neutron with lower energy. However, cross-sections for such reactions are very small.

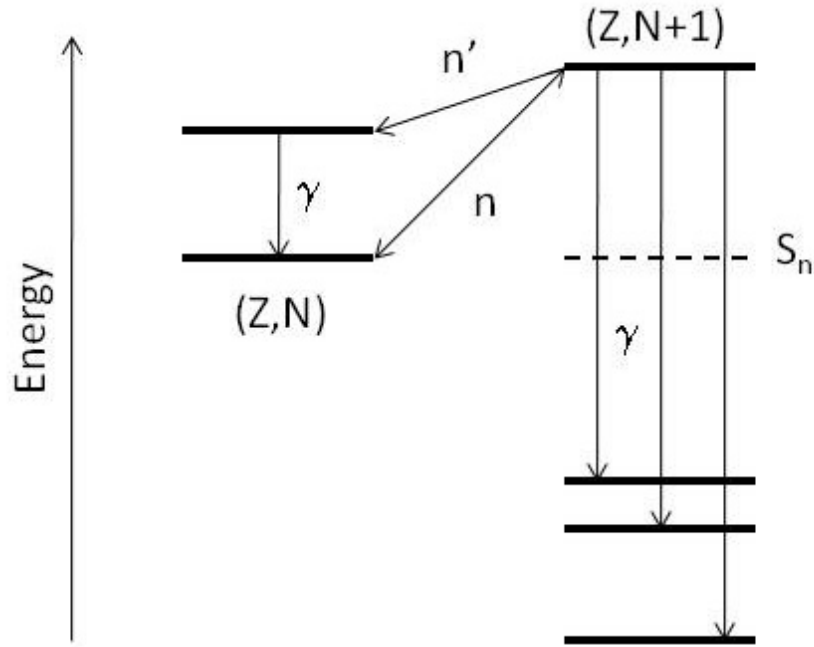

**Figure 3: Schematic view of the (n,n), (n,γ) and (n,n') reactions. The target nucleus is marked as (Z,N) and energy is imagined to increase along the vertical axis. Hence the target nucleus is plotted at the same height as the separation energy $S_n$ for a neutron from the (Z,N+1) nucleus. The arrows indicate some possible reaction paths.**

Figure 4 shows cross section for the $^{14}$N(n,p) reaction between 0.4 and 6.6 MeV. While, again, the 1/v behavior is observed at energies below 10 keV, here, some structure can be seen. These resonances reflect the level structure in the $^{15}$N$^*$ compound nucleus, i.e., the cross section increases in the vicinity of excited states in the compound system. From tabulated values of binding energies one gets $S_n(^{15}N)$ = 10.835 MeV and $S_p(^{15}N)$ = 10.208 MeV. Hence, we obtain Q($^{14}$N(n,p)) = $S_n(^{15}N)$ - $S_p(^{15}N)$ = 0.626 MeV. Adding $S_n(^{15}N)$ to the incoming neutron energy we can see that, e.g., the large peak roughly in the middle of Fig. 4 at about 1.4 MeV reflects the 12.15 MeV state in $^{15}$N.

With increasing energy so-called multiple compound emission becomes possible and reactions such as (n,np) and (n,2n) occur. A nice illustration is provided in Fig. 5 showing multiple chance fission of $^{238}$U. Note how the total fission cross section increases in distinct steps. The threshold for $^{238}$U(n,fission) is at about 1 MeV. Just above this threshold, the only possible system that actually fissions is $^{239}$U$^*$, which is so-called first chance fission. With increasing energies it becomes possible that the original compound nucleus emits an increasing number of neutrons before fission takes place. Besides first chance fission, Fig. 5 also shows second and third chance fission in which the fissioning nuclei are $^{238}$U$^*$ and $^{237}$U$^*$, respectively.

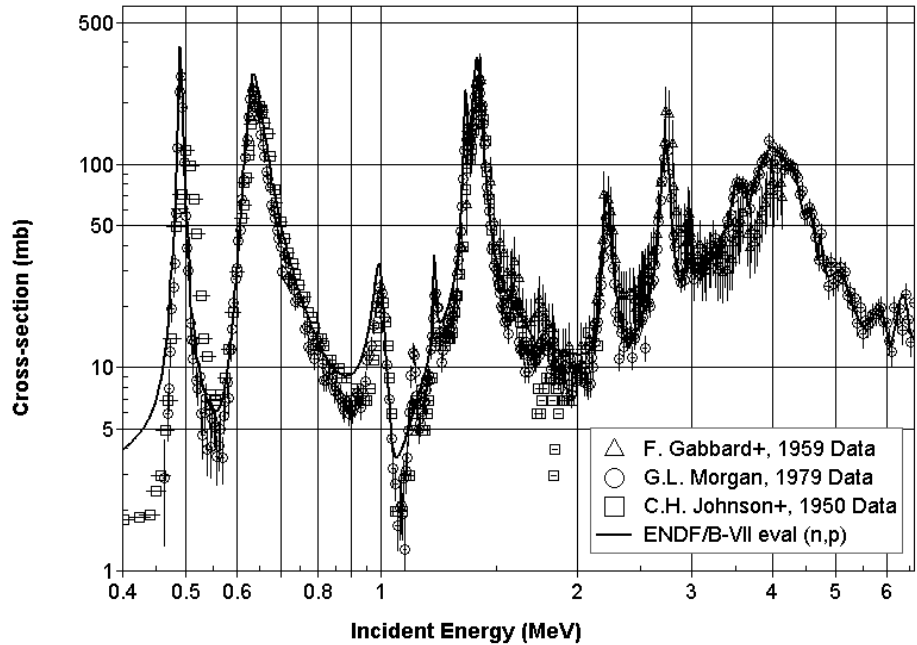

Figure 4: Cross section for the $^{14}N(n,p)^{14}C$ reaction plotted as function of the incident neutron energy. The observed structure carries information about excited states in the $^{15}N^*$ compound nucleus. The solid line shows the ENDF/B-VII evaluation, the other data points are from experimental observations as given in the EXFOR database (EXFOR; Henriksson et al., 2008). All data were retrieved with Janis 3.0.

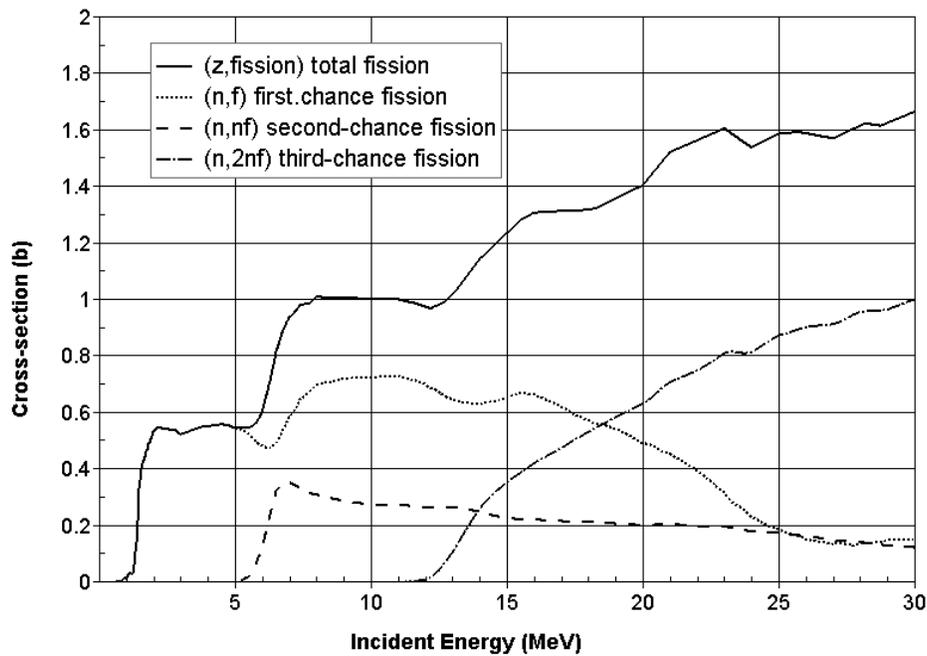

Figure 5: Cross section for the $^{238}U(n,fission)$ reaction plotted as function of the incident neutron energy (solid). At first only first chance fission is possible (dotted). Onset of second (dashed) and third (dashed-dotted) chance fission occur at about 5 and 12 MeV. The data are from the ENDF/B-VII evaluation.

At still higher incoming neutron energies pre-equilibrium processes, direct reactions and intra-nuclear cascade processes become important in nuclear reactions. These processes or models consider that the incoming particle carries information about its original energy and direction during the first few interactions with nucleons of the target nucleus. These interactions may lead to further fast particles inside the nucleus which might be viewed as strongly excited particle-hole pairs. Gradually, equilibrium is reached and the compound state is formed. This is illustrated in Fig. 6 which schematically shows the energy spectrum of emitted particles. At the high-energy end direct reactions dominate. Only one or two interactions between the incoming neutron and nucleons in the nucleus have taken place and nuclear structure effects show up as discrete peaks. With increasing reaction time and number of neutron-nucleon interactions, first the pre-equilibrium and finally the compound domain is reached.

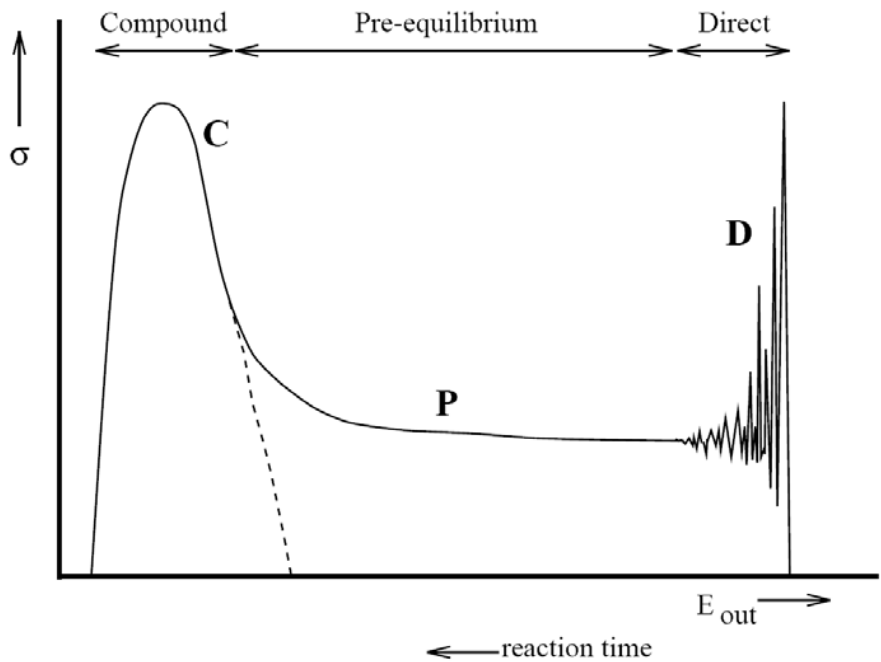

**Figure 6: Schematic drawing of of an outgoing particle energy spectrum. The energy regions to which direct (D), pre-equilibrium (P) and compound (C) mechanism contribute are indicated. The dashed curve distinguishes the compound contribution from the rest in the transitional energy region. The figure is taken from the TALYS user manual (Koning et al., 2007).**

Figure 7 shows double-differential cross section data for $^{nat}Si(n,px)$ at 96 MeV (Tippawan et al., 2004). The cross-section at the high-energy end of the energy spectra of the emitted protons decreases rapidly with the angle. For low proton-emission energies the cross-section remains almost constant, i.e., emission is almost isotropic and independent of the incoming neutron direction.

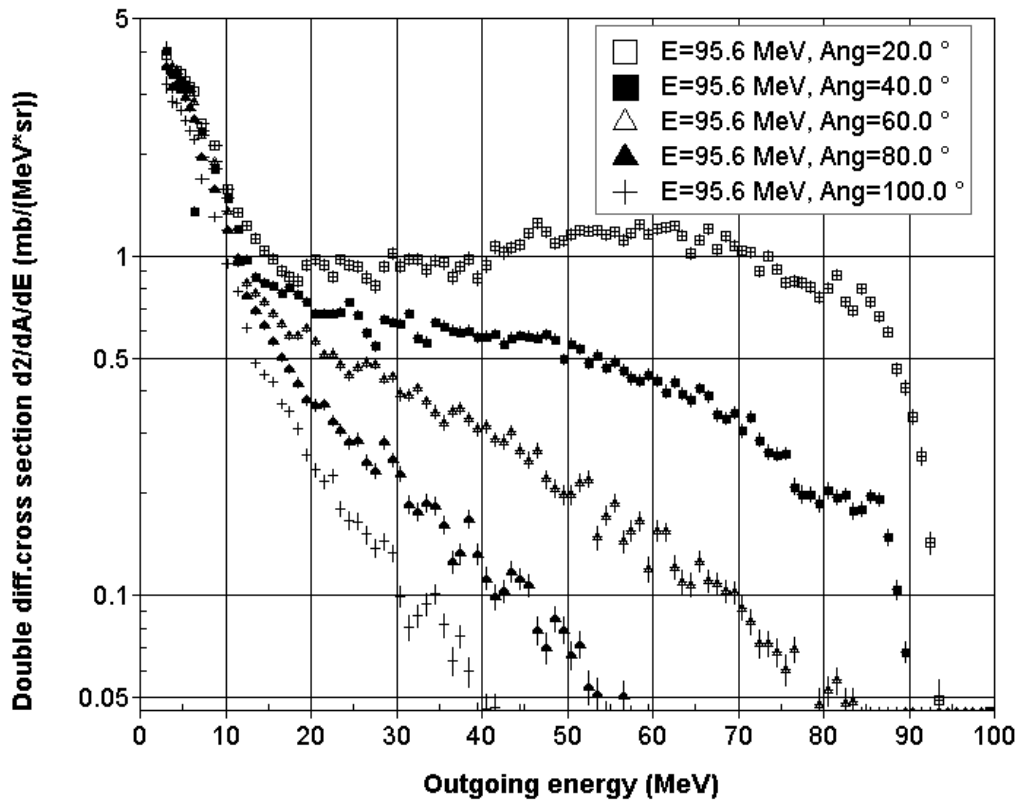

Figure 7: Double-differential cross section for the $^{nat}$Si(n,px) at 96 MeV for 5 different emission angles (20°, 40°, 60°, 80° and 100°) as function of the energy of the outgoing proton. The cross section at high proton energies decreases rapidly with increasing emission angle. Data are from Tippawan et al. (2004) and have been retrieved from the EXFOR database with Janis 3.0 (cp. Fig. 4).

A comprehensive and easy to use nuclear model code is the TALYS code (Koning et al., 2008). One particular advantage of this code is the existence of an extensive manual (Koning et al., 2007). TALYS selects by itself the models that are best suited to describe a certain reaction in a certain energy domain. The readable ascii output can in an easy way be used to, e.g., study the varying contributions of compound, pre-equilibrium and direct reactions.

**Which data are relevant?**

Since the human body consists to about 95% of hydrogen, carbon and oxygen (ICRU Report 63, 2000), cross-section data for neutron interaction with these nuclei are the most important for determination of the absorbed dose in tissue. Other important nuclei are nitrogen and calcium. Roughly half the dose due to neutrons of several tens of MeV comes from proton recoils, i.e., elastic scattering on hydrogen. The remaining part comes to about 10-15% from nuclear recoils due to elastic neutron scattering and to 35-40% from neutron-induced production of light ions, i.e., protons, deuterons, tritons, $^3$He and alpha particles (Blomgren and Olsson, 2003). While the data situation for neutron interaction with hydrogen is rather good, sufficiently accurate data on light-ion production from carbon, nitrogen, oxygen, and calcium are still needed (Chadwick et al., 1997). Some progress has recently been made for carbon and oxygen where double-differential cross sections at 96 MeV have been measured (Tippawan et al., 2006, 2009a, 2009b). Such data are then used to guide and validate evaluations which combine experimental and theoretical information into complete data sets of double-differential cross sections (see, e.g., ICRU Report 63 (2000) and Chadwick et al. (1999)). With these, one can, e.g., accurately calculate dose distributions in radiotherapy.

Besides these needs for dose determination in the human body, other data are of importance for improving dosimetry. These include data on nuclear reactions used or responsible for neutron production and neutron detection. Such data are needed to achieve better accuracy in the description of the radiation environment, e.g., in airplanes or at the treatment place (see, e.g., Bartlett (2009)).

**Neutron sources and beams**

There are several common neutron sources or generators for energies up to about 20 MeV. Several of them combine use an alpha source together with ($\alpha$,n) reactions. Examples are Am-Be sources using the $^9$Be($\alpha$,n) reaction with a Q-value of 5.7 MeV and Am-Li sources (Lebreton et al., 2009; Tsujimura and Yoshida, 2009; and several other contributions in these proceedings). Neutron fields are also commonly created using DD and DT generators employing the D(d,n)$^3$He and T(d,n)$^4$He reactions. A useful reaction which can produce mono-energetic neutrons at various energies in the keV domain is $^{45}$Sc(p,n) (Tanimura et al., 2007; Lamirand et al., 2009). An overview of radiation sources and calibration facilities is given by Lacoste (2009).

A means of using white neutron spectra but allowing for accurate determination of the incoming neutron energy by time-of-flight measurements is offered by the GELINA facility in Geel, Belgium (Bensussan et al., 1978). This facility uses an electron beam hitting a uranium target. The resulting Bremsstrahlung produces neutrons via the $^{238}$U($\gamma$,n), $^{238}$U($\gamma$,2n), and $^{238}$U($\gamma$,fission) reactions. Due to the time structure of the electron beam, the neutron pulses are very short in time, about 1 ns. The resulting resolution for neutron energies in the keV range and at the maximum distance of 400 m reaches about $10^{-3}$.

Several of the above sources and reactions can give truly mono-energetic neutrons. At higher energies this is no longer possible and neutron spectra are either quasi-monoenergetic or "white". Quasi-monoenergetic beams are often produced using the $^7$Li(p,n) reaction. The neutron spectra consist of a peak close to energy of the incoming proton and a broad and roughly even distribution down to zero energy. An example from The Svedberg Laboratory (TSL) is shown in Fig. 8 (Pomp et al., 2005). Each of these components contains about half the neutron intensity. The width of the peak in the figure is, due to experimental resolution, wider than the true width. The latter is determined by the energy loss of the proton when passing the

Li target, which is 4 mm in the shown case, and by the fact that the $^7$Li(p,n)$^7$Be reaction (Q = -1.6 MeV) reaches both the ground state of $^7$Be and the first excited state at 0.43 MeV. The resulting width of the square distribution of the neutron peak is about 2.7 MeV for the shown case and the peak position is at 46.7 MeV. Several facilities offer such beams besides TSL, e.g., iTemba Labs and several laboratories in Japan (Pomp et al. 2005; Nolte et al., 2004; Harano, 2009).

The highest neutron energies available at facilities for nuclear physics experiments are produced from spallation reactions. High-energy proton beams impinging on targets such as tungsten and lead produce white neutron beams with energies up to several hundreds of MeV. Facilities offering such beams are, e.g., n_TOF at CERN (Borcea et al., 2003), LANSCE at Los Alamos (Rochman et al., 2004), GNEIS in Gatchina, Russia (Abrosimov et al., 1985), and, at lower energies, TSL (Prokofiev et al., 2009).

Besides offering the possibility to measure at high neutron energies, the advantages of such facilities is that the neutron energy spectra are similar to atmospheric neutron spectra. This allows, e.g., for so-called accelerated testing of electronics in a neutron beam resembling the environmental neutron field (Prokofiev et al., 2009; Dyer et al., 2009).

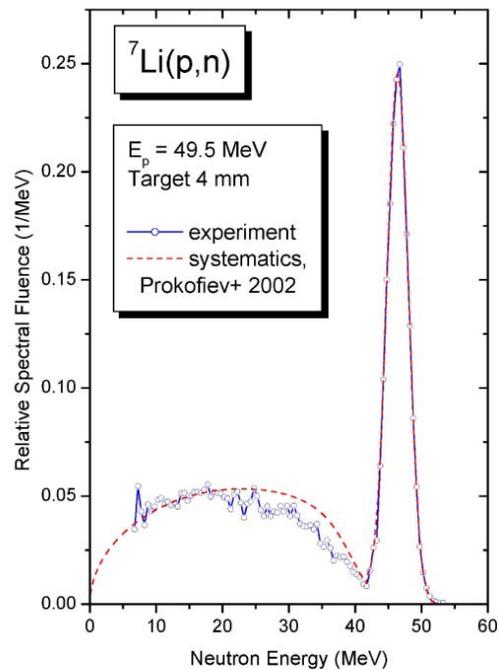

Figure 8: Neutron spectrum from the 7Li(p,n) reaction as measured at TSL for an incoming proton energy of 49.5 MeV. The width of the peak in this measurement is mainly due to the experimental resolution. The systematic (dashed line) has been broadened to reflect this. The used Li target thickness in this case was 4 mm (Pomp et al., 2005).

**A sample experiment**

The Medley experiment at TSL measures neutron-induced double-differential cross section for light ion production. Measurements for a variety of targets have been performed at 96 MeV (Tippawan et al., 2004, 2006, 2009a, 2009b; Blideanu et al., 2004; Bevilacqua et al., 2009). Currently a similar campaign at 175 MeV is under way (Andersson et al., 2009). The light charged-particles emitted in the interaction of the quasi-monoenergetic neutrons with the target nuclei are registered event-by-event in eight telescopes placed at different angles. The telescope consists of two silicon surface barrier detectors of different thickness and a several cm long CsI crystal at the end. With this arrangement the $\Delta$E-E technique can be used for identification of protons, deuterons, tritons, $^3$He and alpha particles. Events due to peak neutrons are selected using time-of-flight information. Fig. 7 shows some typical measured spectra.

A cross section data set for one target nuclei consists of several hundred data points; five kinds of particles are measured at eight different angles and with energies ranging from 2 MeV up to the region of the incoming neutron energy. It is clear that a huge amount of information is lost if these data are to be condensed into kerma coefficients in the way described above. Results for kerma coefficients and kerma ratios obtained from Medley data are given by Göttsche et al. (2009) and Tippawan et al. (2009b).

**Conclusion**

Neutron physics has been researched for many decades now and although a lot of progress has been made much is still left to be done. Due to data needs for dosimetry, especially radiation effects on electronics and humans in space, and the worldwide renewed interest in nuclear power the field is alive and well and will offer excellent research challenges and opportunities even during the next decades.

**Acknowledgements**

This work was supported by the Swedish Natural Science Research Council, the Swedish Cancer Foundation, the Swedish Nuclear Fuel and Waste Management Company, the Swedish Nuclear Power Inspectorate, Ringhals AB, and the Swedish Defence Research Agency.

## References

Abrosimov, N.K., Borukhovich, G.Z., Laptev, A.B., Petrov, G.A., Marchenkov, V.V., Shcherbakov, O.A., Tuboltsev, Yu.V. and Yurchenko, V.I., 1985. *Neutron time-of-flight spectrometer GNEIS at the Gatchina 1 GeV proton synchrocyclotron*. Nucl. Instr. and Meth. A **242**, 121.

Andersson, P. and 12 others, 2009. *Measurements of elastic neutron scattering at 175 MeV*. These proceedings.

Bartlett, D., 2009. *Occupational exposure to cosmic radiation*. These proceedings.

Bensussan, A. and Salome, J. M., 1978. *GELINA: a modern accelerator for high-resolution neutron time of flight experiments*. Nucl. Instrum. Methods **155**, 11.

Bevilacqua, R. and 14 others, 2009. *Neutron-induced light-ion production from iron and bismuth at 175 MeV*. These proceedings.

Blideanu, V. and 34 others, 2004. *Nucleon-induced reactions at intermediate energies: New data at 96 MeV and theoretical status*. Phys. Rev. C **70**, 014607.

Blomgren, J., Olsson, N., 2003. *Beyond kerma – Neutron data for biomedical applications*, Radiat. Prot. Dosim. **103**, 293.

Borcea, C. and 17 others, 2003. *Results from the commissioning of the n_TOF spallation neutron source at CERN*. Nucl. Instr. and Meth. A **513**, 524.

Chadwick, M.B., DeLuca, P.M., Jr and Haight, R.C., 1997. *Nuclear data needs for neutron therapy and radiation protection*, Radiat. Prot. Dosim. **70**, 1.

Chadwick, M.B. and 11 others, 1999. *A consistent set of neutron kerma coefficients from thermal to 150 MeV for biologically important materials*. Med. Phys. **26**, 974.

Dyer, C.S., Truscott, P.R. and Lei, F., 2009. *Overview of radiation damage to electronics*. These proceedings.

EXFOR. http://www.nndc.bnl.gov/exfor/exfor00.htm.

Göttsche, M., Pomp, S., Tippawan, U., Andersson, P., Bevilacqua, R., Blomgren, J., Gustavsson, C., Österlund, M. and Simutkin, V., 2009. *C/O kerma coefficient ratio for 96 MeV neutrons deduced from microscopic measurements*. These proceedings.

Harano, H., 2009. *Monoenergetic and quasi-monoenergetic neutron reference fields in Japan*. These proceedings.

Henriksson, H. Schwerer, O., Rochman, D. Mikhaylyukova, M.V. and Otuka, N., 2008. The art of collecting experimental data internationally: EXFOR, CINDA and the NRDC network. Proceedings of the International Conference on Nuclear Data for Science and Technology, April 22-27, 2007, Nice, France, edited by O. Bersillon, F. Gunsing, E. Bauge, R. Jacqmin, and S. Leray, (EDP Sciences), pp. 737-740.

ICRU Report **63**, 2000. *Nuclear data for neutron and proton radiotherapy and for radiation protection*. (Bethesda, MD: ICRU).

Koning, A.J., Hilaire, S. and Duijvestijn, M.C., 2007. *TALYS 1.0 User Manual*. The TALYS code (see Koning et al. (2008)) and the manual are available for download at http://www.talys.eu.

Koning, A.J., Hilaire, S. and Duijvestijn, M.C., 2008. *TALYS-1.0*. Proceedings of the International Conference on Nuclear Data for Science and Technology, April 22-27, 2007, Nice, France, edited by O. Bersillon, F. Gunsing, E. Bauge, R. Jacqmin, and S. Leray, (EDP Sciences), pp. 211-214.

Krane, K.S., 1988. *Introductory nuclear physics*, John Wiley & Sons, New York.

Lacoste, V., 2009. *Review of radiation sources, calibration facilities and simulated workplace fields*. These proceedings.

Lamirand, V., Gressier, V. and Liatrad, E., 2009. *Comparison of nuclear reactions for the production of monoenergetic neutron fields with energies below 100 keV*. These proceedings.

Lebreton, L., Zimbal, A. and Thomas, D., 2007. *Experimental comparisons of $^{241}$Am – Be neutron fluence energy distributions*. Radiat. Prot. Dosim. **126**, 3.

Nolte, R. and 12 others, 2004. *Quasi-monoenergetic neutron reference fields in the energy range from thermal to 200 MeV*. Radiat. Prot. Dosim. **110**, 97.

Pomp, S. and 17 others, 2005. *The new Uppsala neutron beam facility*. Proceedings of International Conference on Nuclear Data for Science and Technology, Santa Fé, NM, September 26 - October 1 2004, AIP Conference Proceedings No. 769 (Melville, New York), pp. 780-783.

Prokofiev, A. V. , Blomgren, J., Majerle, M., Nolte, R., Röttger, S., Platt, S. P. and Smirnov, S. P., 2009. *Characterization of the ANITA neutron source for accelerated SEE testing at The Svedberg Laboratory*. 2009 IEEE Radiation Effects Data Workshop, Quebec, Canada, July 20-24, pp. 166-173.

Rochman, D., Haight, R.C. , O'Donnell, J.M., Devlin, M., Ethvignot, T. and Granier, T., 2004. *Neutron-induced reaction studies at FIGARO using a spallation source*. Nucl. Instr. and Meth. A **523**, 102.

Soppera, N., Bossant, M., Henriksson, H., Nagel, P. And Rugama, Y., 2008. *Recent upgrades to the nuclear data tool JANIS*. Proceedings of the International Conference on Nuclear Data for Science and Technology, April 22-27, 2007, Nice, France, edited by O. Bersillon, F. Gunsing, E. Bauge, R. Jacqmin, and S. Leray, (EDP Sciences), pp. 773-776. The program is available at http://www.nea.fr/janis/.

Tanimura, Y., Saegusa, J., Shikaze, Y., Tsutsumi, M., Shimizu, S. and Yoshizawa, M., 2007. *Construction of monoenergetic neutron calibration fields using $^{45}$Sc(p,n)$^{45}$Ti reaction at JAEA*. Radiat. Prot. Dosim. **126**, 8.

Tippawan, U. and 20 others, 2004. *Light-ion production in the interaction of 96 MeV neutrons with silicon*. Phys. Rev. C **69**, 064609.

Tippawan, U. and 16 others, 2006. *Light-ion production in the interaction of 96 MeV neutrons with oxygen*. Phys. Rev. C **73**, 034611.

Tippawan, U. and 15 others, 2009a. *Light-ion production in the interaction of 96 MeV neutrons with carbon*. Phys. Rev. C **79**, 064611.

Tippawan, U. and 14 others, 2009b. *Double-differential cross sections and kerma coefficients for light-charged particles produced by 96 MeV neutrons on carbon*. These proceedings.

Tsujimura, N. and Yoshida, T., 2009. *Design of a graphite moderated $^{241}$Am – Li neutron field to simulate reactor spectra*. These proceedings.

Tubiana, M., Dutreix, J. And Wambersie, A., 1990. *Introduction to radiobiology*, Taylor & Francis, London.